\documentclass[aps,a4paper,superscriptaddress,twocolumn]{revtex4}
\usepackage{tensor}
\usepackage{graphicx}
\usepackage{amsmath}
\usepackage{amssymb}
\usepackage{enumerate}
\usepackage{subfigure}
\usepackage{tabularx}
\usepackage[colorlinks=true, pdfstartview=FitV, linkcolor=blue, citecolor=red, urlcolor=black, breaklinks=true]{hyperref}
\newcommand{\be}{\begin{equation}}
\newcommand{\ee}{\end{equation}}
\newcommand{\ben}{\begin{eqnarray}}
\newcommand{\een}{\end{eqnarray}}
\newcommand{\bes}{\begin{subequations}}
\newcommand{\ees}{\end{subequations}}
\def\bal#1\eal{\begin{align}#1\end{align}}
\newcommand{\vphi}{\varphi}

\newcommand{\LL}{{\mathcal L}}

\newcommand{\A}{\mathcal{A}}
\newcommand{\D}{\mathcal{D}}
\newcommand{\F}{\mathcal{F}}

\begin{document}
\title{Vortices in a generalized Maxwell-Higgs model with visible and hidden sectors}
\author{D. Bazeia}\affiliation{Departamento de F\'\i sica, Universidade Federal da Para\'\i ba, 58051-970 Jo\~ao Pessoa, PB, Brazil}
\author{L. Losano}\affiliation{Departamento de F\'\i sica, Universidade Federal da Para\'\i ba, 58051-970 Jo\~ao Pessoa, PB, Brazil}
\author{M.A. Marques}\affiliation{Departamento de F\'\i sica, Universidade Federal da Para\'\i ba, 58051-970 Jo\~ao Pessoa, PB, Brazil}
\author{R. Menezes}\affiliation{Departamento de Ci\^encias Exatas, Universidade Federal da Para\'{\i}ba, 58297-000 Rio Tinto, PB, Brazil}\affiliation{Departamento de F\'\i sica, Universidade Federal da Para\'\i ba, 58051-970 Jo\~ao Pessoa, PB, Brazil}
\begin{abstract}
We investigate the presence of vortices in generalized Maxwell-Higgs models with a hidden sector. The model engenders $U(1)\times U(1)$ symmetry, in a manner that the sectors are coupled via the visible magnetic permeability depending only on the hidden scalar field. We develop a first order framework in which the hidden sector decouples from the visible one. We illustrate the results with two specific examples, that give rise to the presence of vortices with internal structure.
\end{abstract}

\date{\today}
\maketitle

\section{Introduction}
Topological structures appear in high energy physics in a variety of dimensions \cite{B1,B2,weinberg}. Among the mostly known ones, there are kinks, vortices and monopoles, which are static configurations that appear in $(1,1)$, $(2,1)$ and $(3,1)$ spacetime dimensions, respectively. These structures have been investigated for more than 40 years, and they find a myriad of applications in high energy and condensed matter physics \cite{B1,B2,weinberg,fradkin}.

Vortices, in particular, were firstly investigated by Helmholtz in Ref.~\cite{helmholtz} and are commonly found in fluid mechanics \cite{fluidmec}. They also appear in condenser matter, in the study of type II superconductors under specific conditions \cite{abrikosov} when one deals with the Ginzburg-Landau theory of superconductivity \cite{glvortex}. As it is known, the Ginzburg-Landau theory is non-relativistic and the  first relativistic model that supports vortex configurations was introduced by Nielsen and Olesen in Ref.~\cite{NO}; see also Refs.~\cite{vega,bogo}. It consists of a Higgs field coupled to a gauge field with the Maxwell dynamics under a local $U(1)$ symmetry. These vortices are electrically neutral but have quantized magnetic flux.

They were largely investigated in several different contexts, including the case of generalized models; see Refs.~\cite{g0,g1,g2,g3,g4,g5,g6,g7,g8,vv1,vv2,vv3,vv4,vv5,vv6,vtwin,compvortex,godvortex,anavortex}. In Refs.~\cite{g0,g1}, for instance, the presence of a generalized magnetic permeability in a model of the Maxwell-Higgs type allowed to simulate properties of Chern-Simons vortices \cite{csjackiw}. By using a similar model in Ref.~\cite{compvortex}, we have found the presence of compact vortices, which seems to map the magnetic field of an infinitely long solenoid. Due to the aforementioned works with generalized kinematics, we have introduced a first order formalism to describe vortices in generalized models in Ref.~\cite{godvortex} and a procedure to find analytic solutions in Ref.~\cite{anavortex}.

More recently, in Ref.~\cite{intvortex}, we have enlarged the $U(1)$ symmetry to accommodate a $Z_2$ symmetry, with the addition of a single neutral scalar field that interacts with the gauge field via the magnetic permeability. In this case, it was shown that the neutral field acts as a source to generate the vortex. In the past, a similar idea was developed, but with an $U(1)\times U(1)$ symmetry engendered by two gauge fields and two complex scalar fields interacting through an extension of the standard Higgs-like potential \cite{witten}. This model is of interest in the study of superconducting strings, and has been used in several contexts, in particular in Refs.~\cite{sup1,sup2,sup3} and in references therein. For instance, in Ref.~\cite{sup2} the authors describe the presence of novel solutions for non-Abelian cosmic strings and in \cite{sup3} the study focuses on the critical behavior of magnetic field field in a superconductor coupled to a superfluid, intending to simulate the core of neutrons stars, where superconducting protons are supposed to couple with superfluid neutrons. Among other possibilities, the $U(1)$ symmetry may be enlarged to accommodate other fields. In particular, in Refs.~\cite{shif1,shif2,shif3} the $U(1)$ symmetry is enlarged to become $U(1)\times SO(3)$, in which the $SO(3)$ symmetry is governed by the addition of some neutral scalar fields. Furthermore, in the recent work \cite{mono} we considered the symmetry $SU(2)\times Z_2$, extending the study of monopoles in the non Abelian model considered Refs.~\cite{tH,P} to include a new neutral scalar field to produce magnetic monopoles with internal structure in the three-dimensional space.

Models with the $U(1)\times U(1)$ symmetry are useful to include the so-called hidden sector, which seems to be of current interest \cite{ds1,ds2} and may play a role in the study of dark matter \cite{p1,p2,p3,p4}. This hidden (or dark) sector may be coupled to the visible one via the Higgs fields, as it appears in the Higgs portal \cite{hp1,hp2}, in a way similar to the coupling that appeared before in Refs.~\cite{witten,sup1,sup2}. Another possibility is to add the interaction of the gauge field strengths as in \cite{dm1,dm2,dm3,dm4,dm5}, with the two strength tensors coupled to each other. We can also enlarge the symmetry and consider $SU(2)\times U(1)$, as in \cite{sup2}, for instance, 
and also $SU(3)\times U(1)$, as investigated very recently in \cite{cm}, to search for color-magnetic structures in dense quark matter, compatible with the interior of compact stars and able to produce detectable gravitational waves. In this work, however, we follow a different direction and consider the $U(1)\times U(1)$ symmetry, choosing the magnetic permeability of the visible sector to be driven by the scalar field of the hidden sector, without the coupling between the electromagnetic strength tensors. This is a novel possibility, which leads to results of current interest. 

There are other interesting issues that suggest the study of topological structures in models described by the $U(1)\times U(1)$ symmetry, as in \cite{dm4,dm5} and in references therein. In this work, however, we deal with a relativistic model in $(2,1)$ spacetime dimension and, inspired by Refs.~\cite{NO,vega,bogo}, we consider a model in which the first $U(1)$ portion describes the visible sector and the second accounts for the hidden sector. The system engenders two complex scalars of the Higgs type, the visible $\vphi$ and the hidden $\chi$ fields, with the coupling between the two sectors coming from a nonnegative function $P(|\chi|)$ that is controlled by the hidden scalar field. As we show below, the function $P(|\chi|)$ describes the magnetic permeability of the visible sector, and the main motivation for this is the fact that the dark matter seems to be always permeating the medium where the visible or baryonic matter appears to form galaxies in the Universe. Although we are dealing with a three-dimensional toy model, we believe that it is worth exploring this kind of coupling, since we still have no important insight on how the visible matter interacts with the dark component. Also, we will be searching for static vortex-like configurations at the classical level, intending to see how the topological structure of the hidden sector may modify the profile of the corresponding visible counterpart.

Dark matter is one among several problems that cannot be solved inside the Standard Model of particle physics, so the enhancement of the $U(1)$ symmetry to the case $U(1)\times U(1)$ is welcome to the study involving the visible and hidden sectors; see, e.g., the recent works \cite{DM1,DM2} and references therein, on several open problems in physics, in particular on dark matter. Moreover, since we are working at the classical level, the current study makes no progress on possible quantum features of the model. However, it can also be of interest to study pipelike vortices in two-component condensates \cite{PL} and also, the interaction between vortices, that can find applications in magnetic materials, for instance, if one changes the formulation in terms of the visible and hidden sectors to the case of hard and soft arrangements that appear in bimagnetic materials; see, e.g., \cite{BM,BIM} and references therein. In this sense, the doubling of the $U(1)$ symmetry to the $U(1\times U(1))$ case is of good use to investigate systems composed of two components condensates, not only in the case of visible and hidden sectors \cite{dm1,dm2,dm3,dm4}, but also in other contexts, such as the normal and superconductor components firstly studied in \cite{witten}, the superconductor and superfluid coupling that is supposed to be present in the core of neutron stars \cite{sup3}, the study of pipelike vortex in two-components condensates \cite{PL}, and the possibility of building two-component magnetic materials, with a hard magnetic component interacting with a soft magnetic component of the bimagnetic material \cite{BM,BIM}.

In order to introduce and investigate the $U(1)\times U(1)$ model, we organize the paper as follows: in Sec.~\ref{secmodel}, we present the model and develop the Bogomol'nyi procedure \cite{bogo}, to find first order differential equations which are very relevant to describe stable vortex configurations. For the proposed model, we also show that the first order framework acts to decouple the hidden sector from the visible one. This is an interesting result, and we can then use the hidden sector as a source to describe the visible sector. In Sec.~\ref{secex}, we illustrate the  results with two examples, which give rise to vortices with features of current interest. We end the work in Sec.~\ref{secconc}, where we add conclusions and perspectives for future works.

\section{The general model}\label{secmodel}
We work in $(2,1)$ flat spacetime dimensions with the action $S=\int d^3x \LL$, where the Lagrangian density is
\be\label{lmodel}
\begin{aligned}
	\LL &= - \frac{1}{4}P(|\chi|)F_{\mu\nu}F^{\mu\nu} - \frac{1}{4}Q(|\chi|)\F_{\mu\nu}\F^{\mu\nu} \\
	    &\hspace{4mm} + |D_\mu\vphi|^2 + |\D_\mu\chi|^2 - V(|\vphi|,|\chi|).
\end{aligned}
\ee
Here $\vphi$ and $A_\mu$ are the visible complex scalar and gauge fields, and the corresponding hidden fields are $\chi$ and $\A_\mu$. As usual, $F_{\mu\nu}=\partial_{\mu}A_\nu-\partial_{\nu}A_\mu$ is the visible electromagnetic strength tensor, and $D_\mu=\partial_{\mu}+ieA_{\mu}$ stands for the covariant derivative. Their equivalent hidden counterparts are $\F_{\mu\nu}=\partial_\mu \A_\nu-\partial_\nu \A_\mu$ and $\D_\mu=\partial_{\mu}+iq\A_{\mu}$. The potential is denoted by $V(|\vphi|,|\chi|)$ and may present terms that couple both the visible and hidden scalar fields. In principle, the explicit form of the potential is unknown, and the generalized magnetic permeabilities are controlled by
$P(|\chi|)$ and $Q(|\chi|)$, which are non negative functions that depend exclusively on the field $\chi$. The above gauge invariant Lagrangian density engenders the $U(1)\times U(1)$ symmetry. The inclusion of functions of the scalar field multiplying the Maxwell term is also present in models related to holography; see, e.g., \cite{HI,roge} for two distinct possibilities, used to describe an holographic insulator model with nonsingular zero temperature infrared geometry \cite{HI}, and the electric charge transport in a strongly coupled quark-gluon plasma \cite{roge}.

In our model, the visible and hidden sectors are coupled through $P(|\chi|)$, i.e., the hidden scalar field controls the magnetic permeability of the visible sector. This is a different approach from the one considered in Ref.~\cite{dm5}, where the coupling was done with the electromagnetic strength tensors, through the term $F_{\mu\nu}\F^{\mu\nu}$, which we are not considering in the current work. Here, we use the metric tensor
$\eta_{\mu\nu}=(1,-1,-1)$ and the natural units $\hbar=c=1$.

By varying the action with respect to the fields, we get that the equations of motion associated to the Lagrangian density \eqref{lmodel} are
\bes\label{geom}
\begin{align}
 D_{\mu}D^\mu\vphi &= -\frac{\vphi}{2|\vphi|} V_{|\vphi|}, \\
  \D_{\mu} \D^\mu \chi &= -\frac{\chi}{2|\chi|}\bigg(\! \frac{P_{|\chi|}}{4}F_{\mu\nu}F^{\mu\nu} \!+\! \frac{Q_{|\chi|}}{4}\F_{\mu\nu}\F^{\mu\nu} \!+\! V_{|\chi|}\!\bigg), \\ \label{meqsa}
 \partial_\mu \left(P F^{\mu\nu}\right) &= J^\nu, \\ \label{meqsc}
 \partial_\mu \left(Q \F^{\mu\nu}\right) &= \mathcal{J}^\nu,
\end{align}
\ees
where the currents are $J_{\mu} = ie\left(\overline{\vphi}\,D_{\mu} \vphi-\vphi\,\overline{D_{\mu}\vphi}\right)$ and $\mathcal{J}_{\mu} = iq\left(\overline{\chi}\,\D_{\mu} \chi-\chi\,\overline{\D_{\mu}\chi}\right)$. Also, we have used the notation $V_{|\vphi|} = \partial V/\partial|\vphi|$, $V_{|\chi|} = \partial V/\partial|\chi|$ and so on. By setting $\nu=0$ in Eqs.~\eqref{meqsa} and \eqref{meqsc} and considering static field configurations, one can show the Gauss' laws are identities for $A_0=0$ and for $\A_0=0$. In this case, the solutions are electrically neutral since the electric charges vanish.

Invariance under spacetime translations $x^\mu\to x^\mu + b^\mu$, with $b^\mu$ constant, lead to the conserved energy-momentum tensor
\be\label{emtgeneral}
\begin{aligned}
	T_{\mu\nu} &= P F_{\mu\lambda}\tensor{F}{^\lambda_\nu} + Q \F_{\mu\lambda}\tensor{\F}{^\lambda_\nu} +\overline{D_{\mu} \vphi}D_{\nu} \vphi + \overline{D_{\nu} \vphi}D_{\mu} \vphi \\
	&\hspace{4mm} + \overline{\D_{\mu} \chi}\D_{\nu} \chi + \overline{\D_{\nu} \chi}\D_{\mu} \chi - \eta_{\mu\nu} \LL.
\end{aligned}
\ee
To search for vortexlike solutions, we consider static configurations and the usual ansatz
\be\label{ansatz}
\begin{aligned}
	\vphi &= g(r)e^{in\theta}, & \chi &= h(r)e^{ik\theta},\\
	\vec{A} &= {\frac{\hat{\theta}}{er}[n-a(r)]}, & \vec{\A} &= {\frac{\hat{\theta}}{qr}[k-c(r)]},
\end{aligned}
\ee
in which $n$ and $k$ are nonvanishing integer numbers that control the vorticity of the visible and hidden solutions, respectively. Although this is not the most general situation, the above cylindrically symmetric ansatz for the fields is largely used in the search for vortex solutions in planar systems; see, e.g., Refs.~\cite{NO,vega,bogo,dm4,dm5} and references therein. Also, when $P(|\chi|)=1$ the visible and hidden sectors decouple, and the expressions in Eq.~\eqref{ansatz} lead us back to the standard situation. The functions $a(r)$, $c(r)$, $g(r)$ and $h(r)$ obey the boundary conditions
\bes\label{bc}
\bal
g(0) &= 0, & h(0)&=0, & a(0) &= n, & c(0)&=k, \\
g(\infty) &=v, & h(\infty) &=w, & a(\infty)&=0, & c(\infty)&=0.
\eal
\ees
Here, $v$ and $w$ are parameters that control the asymptotic behavior of the functions $g(r)$ and $h(r)$. Considering Eqs.~\eqref{ansatz}, the visible and hidden magnetic fields are giving by, respectively,
\be\label{B}
B = -F^{12} = -\frac{a^\prime}{er} \quad\text{and}\quad \mathcal{B} = -\F^{12} = -\frac{c^\prime}{qr},
\ee
where the prime stands for the derivative with respect to $r$. By using this, one can show the magnetic fluxes are quantized
\be\label{mflux}
\begin{aligned}
	\Phi_{(B)} &= 2\pi\int rdr B = \frac{2\pi}{e}n, \\
	\Phi_{(\mathcal{B})} &= 2\pi\int rdr \mathcal{B} = \frac{2\pi}{q}k.
\end{aligned}
\ee
These results allow that we introduce two topological currents, one in the visible sector, defined by $J_T^\mu=\varepsilon^{\mu\nu\lambda}\partial_\lambda A_\nu$, and the other in the hidden sector, defined by
${\cal J}_T^\mu=\varepsilon^{\mu\nu\lambda}\partial_\lambda{\cal A}_\nu$; they are conserved and describe conserved charges that are directly identified with the corresponding magnetic fluxes
\be
\begin{aligned}
	Q_T &= 2\pi \int r dr\, J_T^0=\Phi_{(B)}, \\
	{\cal Q}_T &= 2\pi \int r dr\, {\cal J}_T^0=\Phi_{({\cal B})}.
\end{aligned}
\ee
The above results inform us that the suggested solutions are topologically protected against decay into the elementary excitations of the system.

We use Eqs.~\eqref{ansatz} to rewrite the equations of motion \eqref{geom} in the form
\bes\label{secansatz}
\begin{align}
&\frac{1}{r} \left(r g^\prime\right)^\prime -\frac{a^2g}{r^2} - \frac12  V_{|\vphi|} = 0, \\ 
&\frac{1}{r} \left(r h^\prime\right)^\prime\! -\!\frac{c^2h}{r^2}\! -\! \frac12 \left(\! P_{|\chi|}\frac{{a^\prime}^2}{2e^2r^2}\! + Q_{|\chi|}\frac{{c^\prime}^2}{2q^2r^2}\! +\! V_{|\chi|}\!\right)\! = 0, \\
&r\left(P\frac{a^\prime}{er} \right)^\prime - 2eag^2 = 0, \\
&r\left(Q\frac{c^\prime}{qr} \right)^\prime - 2qch^2 = 0.
\end{align}
\ees
Using the above Eqs.~\eqref{ansatz}, the nonvanishing components of energy-momentum tensor \eqref{emtgeneral}  become
\bes
\bal\label{rhoans}
T_{00} &= P \frac{{a^\prime}^2}{2e^2r^2} + Q \frac{{c^\prime}^2}{2q^2r^2} +  {g^\prime}^2 +\frac{a^2g^2}{r^2} +  {h^\prime}^2 + \frac{c^2h^2}{r^2} + V, \\
T_{12} &= \left({g^\prime}^2 - \frac{a^2g^2}{r^2} + {h^\prime}^2 - \frac{c^2h^2}{r^2} \right) \sin(2\theta), \\ 
T_{11} &= P \frac{{a^\prime}^2}{2e^2r^2}+ Q \frac{{c^\prime}^2}{2q^2r^2} + \left({g^\prime}^2+{h^\prime}^2\right)\!\left(2\cos^2\theta-1\right) \nonumber\\
       &\hspace{4mm} +\left(\frac{a^2g^2}{r^2}+\frac{c^2h^2}{r^2}\right)\!\left(2\sin^2\theta -1\right) -V, \\ 
T_{22} &= P \frac{{a^\prime}^2}{2e^2r^2}+ Q \frac{{c^\prime}^2}{2q^2r^2} + \left({g^\prime}^2+{h^\prime}^2\right)\!\left(2\sin^2\theta-1\right) \nonumber\\
       &\hspace{4mm} +\left(\frac{a^2g^2}{r^2}+\frac{c^2h^2}{r^2}\right)\!\left(2\cos^2\theta -1\right) -V.
\eal
\ees
In the above equations, we identify the energy density as $\rho=T_{00}$ and the stress tensor as $T_{ij}$. The equations of motion \eqref{secansatz}  present coupling between the functions and are of second order. In order to simplify the problem, it is of interest to find a first order formalism for the model. This can be done by using Eq.~\eqref{rhoans} to develop the Bogomol'nyi procedure \cite{bogo}. After some algebraic manipulations, we can write
\be
\begin{aligned}
	\rho &= \frac{P}{2} \!\left(\!\frac{a^\prime}{er} + \frac{e\left(v^2-g^2\right) }{P} \!\right)^2 + \frac{Q}{2}\!\left(\!\frac{c^\prime}{qr} + \frac{ q\left(w^2-h^2\right)}{Q} \!\right)^2 \\
       &\hspace{4mm} + \left(g^\prime - \frac{ag}{r} \right)^2+ \left(h^\prime - \frac{ch}{r} \right)^2 \\
	    &\hspace{4mm} + V - \frac{e^2}{2}\frac{\left(v^2-g^2\right)^2}{P} + \frac{q^2}{2} \frac{\left(w^2-h^2\right)^2}{Q} \\
       &\hspace{4mm}- \frac{1}{r} \left(a\left(v^2-g^2\right) + c\left(w^2-h^2\right)\right)^\prime.
\end{aligned}
\ee
As we stated before, the potential is in principle arbitrary. However, motivated by the Bogomol'nyi procedure \cite{bogo}, the choice 
\be\label{potbogo}
V(|\vphi|,|\chi|) =  \frac{e^2}{2} \frac{\left(v^2-|\vphi|^2\right)^2}{P(|\chi|)}  + \frac{q^2}{2} \frac{\left(w^2-|\chi|^2\right)^2}{Q(|\chi|)},
\ee
allows that one writes the energy in the form
\be\label{eq12}
\begin{aligned}
	E &= 2\pi\int_0^\infty r\,dr\,\frac{P}{2}\! \left(\!\frac{a^\prime}{er} + \frac{e\left(v^2-g^2\right) }{P} \!\right)^2 \\
       &\hspace{4mm}+ 2\pi\int_0^\infty r\,dr\, \frac{Q}{2}\!\left(\!\frac{c^\prime}{qr} + \frac{ q\left(w^2-h^2\right)}{Q} \!\right)^2 \\
	    &\hspace{4mm} + 2\pi\int_0^\infty r\,dr\left(\!g^\prime - \frac{ag}{r} \!\right)^2 + 2\pi\int_0^\infty r\,dr\left(\!h^\prime - \frac{ch}{r} \!\right)^2 \\
       &\hspace{4mm} +E_B,
\end{aligned}
\ee
where 
\be\label{eb}
\begin{split}
	E_B &= 2\pi\int_0^\infty dr\,\left(a\left(v^2-g^2\right) + c\left(w^2-h^2\right)\right)^\prime \\
	    &= 2\pi\left(|n|v^2 + |k|w^2\right).
\end{split}
\ee
Although the choice \eqref{potbogo} is not necessary, it leads us to Eq.~\eqref{eq12}, which unveils the interesting possibility: the first four terms are all non-negative, so we see that the energy is bounded to the value $E=E_B$, if the solutions obey the first order equations
\bes\label{fovisible}
\bal
g^\prime &= \frac{ag}{r}, \\
-\frac{a^\prime}{er} &= \frac{e\left(v^2-g^2\right) }{P(h)},
\eal
\ees
for $a(r)$ and $g(r)$ controlling the gauge and scalar fields in the visible sector, and
\bes\label{fohidden}
\bal
h^\prime &= \frac{ch}{r}, \\
-\frac{c^\prime}{qr} &= \frac{q \left(w^2-h^2\right) }{Q(h)},
\eal
\ees
for $c(r)$ and $h(r)$ controlling the gauge and scalar in the hidden sector. Therefore, we have obtained four first order equations to study the problem. One can show they satisfy the equations of motion \eqref{secansatz} with the potential \eqref{potbogo}. An interesting fact is that solutions of the first order equations are endowed with minimum energy, so they are stable according to the classical result of Bogomol'nyi \cite{bogo} and also, they are compatible with the stressless condition, $T_{ij}=0$, which assures stability of the solutions under rescaling; see Refs.~\cite{godvortex,H,D}. Notice that the first order equations \eqref{fohidden} are decoupled from the ones for the visible sector in Eq.~\eqref{fovisible}. Therefore, we firstly solve the hidden sector to see how it modifies the visible one by choosing $P(|\chi|)$.

Before doing that, however, we notice that the presence of the first order equations \eqref{fohidden} and \eqref{fovisible} allows us to write the energy density \eqref{rhoans} in the form $\rho = \rho_{hidden} + \rho_{visible}$, where
\bes
\bal\label{rhohidden}
\rho_{hidden} =  Q(h) \frac{{c^\prime}^2}{q^2r^2} + 2{h^\prime}^2,\\ \label{rhovisible}
\rho_{visible} = P(h) \frac{{a^\prime}^2}{e^2r^2} + 2{g^\prime}^2,
\eal
\ees
indicate the contribution of the visible and hidden sectors, respectively. We also highlight the fact that, from Eq.~\eqref{eb}, the energy of the visible and hidden sectors are independent and fixed, despite the presence of the functions $P(|\chi|)$ and $Q(|\chi|)$. For the visible sector, we have $E_{visible} = 2\pi v^2 |n|$ and for the hidden one, $E_{hidden} = 2\pi w^2 |k|$.

Another feature of the above procedure, which can be seen from the potential in Eq.~\eqref{potbogo} and the first order equations \eqref{fovisible} and \eqref{fohidden}, is the presence of spontaneous symmetry breaking (SSB) in both the visible and hidden sectors. The presence of SSB in the hidden sector is necessary to give rise to vortex in the sector, and the first term in \eqref{potbogo} induces SSB in the visible sector, and is mandatory to generate the first order equations that attain stable minima energy configurations. In particular, for
$P(|\chi|)=1$ and $Q(|\chi|)=1$, one gets to the case of two uncoupled and similar ($e\leftrightarrow q$ and $v\leftrightarrow w$) Maxwell-Higgs models, each one with its own Nielsen-Olesen vortex configuration \cite{glvortex}. To keep the two sectors coupled one has to make $P(|\chi|)$ nontrivial, and this we discuss in the next Section. Moreover, one has to choose $P(|\chi|)$ and $Q(|\chi|)$ carefully, because they have to lead to first order equations \eqref{fovisible} and \eqref{fohidden} that support solutions compatible with the boundary conditions \eqref{bc}.

\section{Some specific models}\label{secex}

Let us now illustrate our findings with two distinct examples. Before doing so, we can simplify the problem by rescaling the fields as
\be
	\vphi \to v \vphi, \quad \chi \to v\chi, \quad A_\mu \to v A_\mu \quad\text{and}\quad  \mathcal{A}_\mu \to v \mathcal{A}_\mu,
\ee
and taking $r \to r/ev$ and $\LL \to e^2v^4 \LL.$ This leads us to work with dimensionless quantities. We also consider $e=1$ and $v=1$ and understand the parameter $q$ and $w$ as the ratio of the respective constants $(q/e)$ and $(w/v)$ of the hidden and visible sectors.

\subsection{First example}
Considering the hidden sector to be controlled by $Q(|\chi|)=1$, its first order equations \eqref{fohidden} take the form
\bes\label{fono}
\bal
h^\prime &= \frac{ch}{r}, \\
-\frac{c^\prime}{qr} &= q \left(w^2-h^2\right),
\eal
\ees
which admits vortexlike solutions of the Nielsen-Olesen type. Unfortunately, the analytical solutions $a(r)$ and $h(r)$ are currently unknown. Therefore, one has to use numerical methods to find them. Since their profiles are well known (see Refs.~\cite{NO,vega}), we will not display them here.

We then go on and investigate how the hidden fields modify the visible sector. We take the choice
\be\label{P1}
P(|\chi|) = \frac{1}{\left(1-\beta|\chi|^2\right)^2},
\ee
where $\beta$ is a non negative, dimensionless parameter. Notice that $\beta$ controls how strongly the hidden scalar field modifies the magnetic permeability of the visible sector. The case $\beta=0$ leads to $P(|\chi|)=1$, which decouples the sectors. The above function makes the potential in Eq.~\eqref{potbogo} to be written as
\be
V(|\vphi|,|\chi|) =  \frac{1}{2}  \!\left(1-|\vphi|^2\right)^2\!\left(1-\beta|\chi|^2\right)^2 \!+\! \frac{q^2}{2}\!\left(w^2-|\chi|^2\right)^2.
\ee
From Eq.~\eqref{fohidden}, we get that the first order equations for the visible sector are
\bes\label{fohiddenno}
\bal
g^\prime &= \frac{ag}{r}, \\
-\frac{a^\prime}{r} &= \left(1-g^2\right)\left(1-\beta h^2\right)^2.
\eal
\ees
It is straightforward to see that the case $\beta=0$ leads to Nielsen-Olesen solutions, which are very similar to the ones of the hidden sector, except for the constants $q$ and $w$. The parameter $\beta$ plays an important role in the model. One can show that $a^\prime(r)$ vanishes for $h^2=1/\beta$. We then expect to see a region in which $a(r)$ is approximately uniform. Since $0\leq h(r)<1$, this feature appears only for $\beta>1$, which is the range that we will consider here.

\begin{figure}[t!]
\centering
\includegraphics[width=5.2cm]{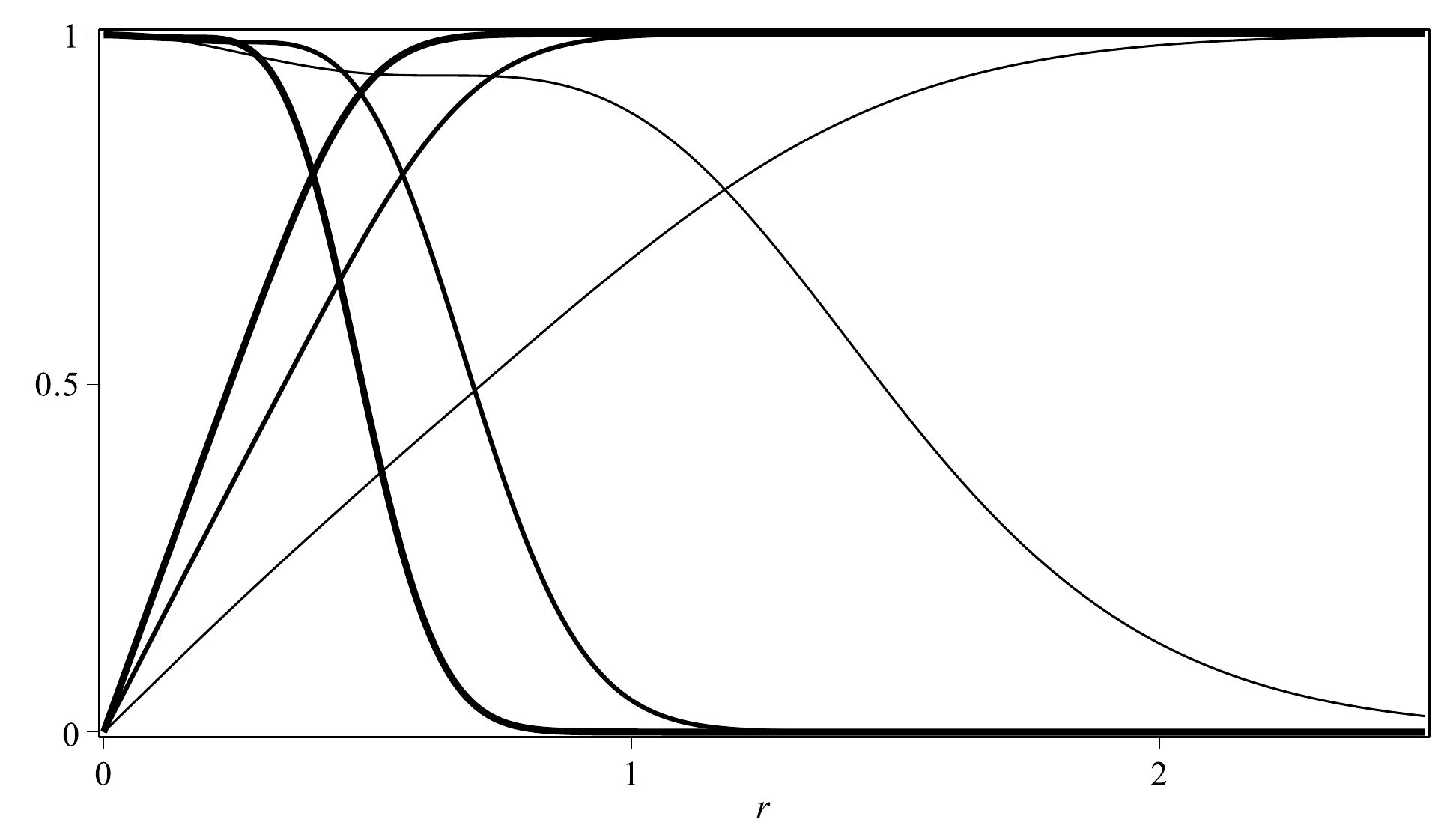}
\includegraphics[width=5.2cm]{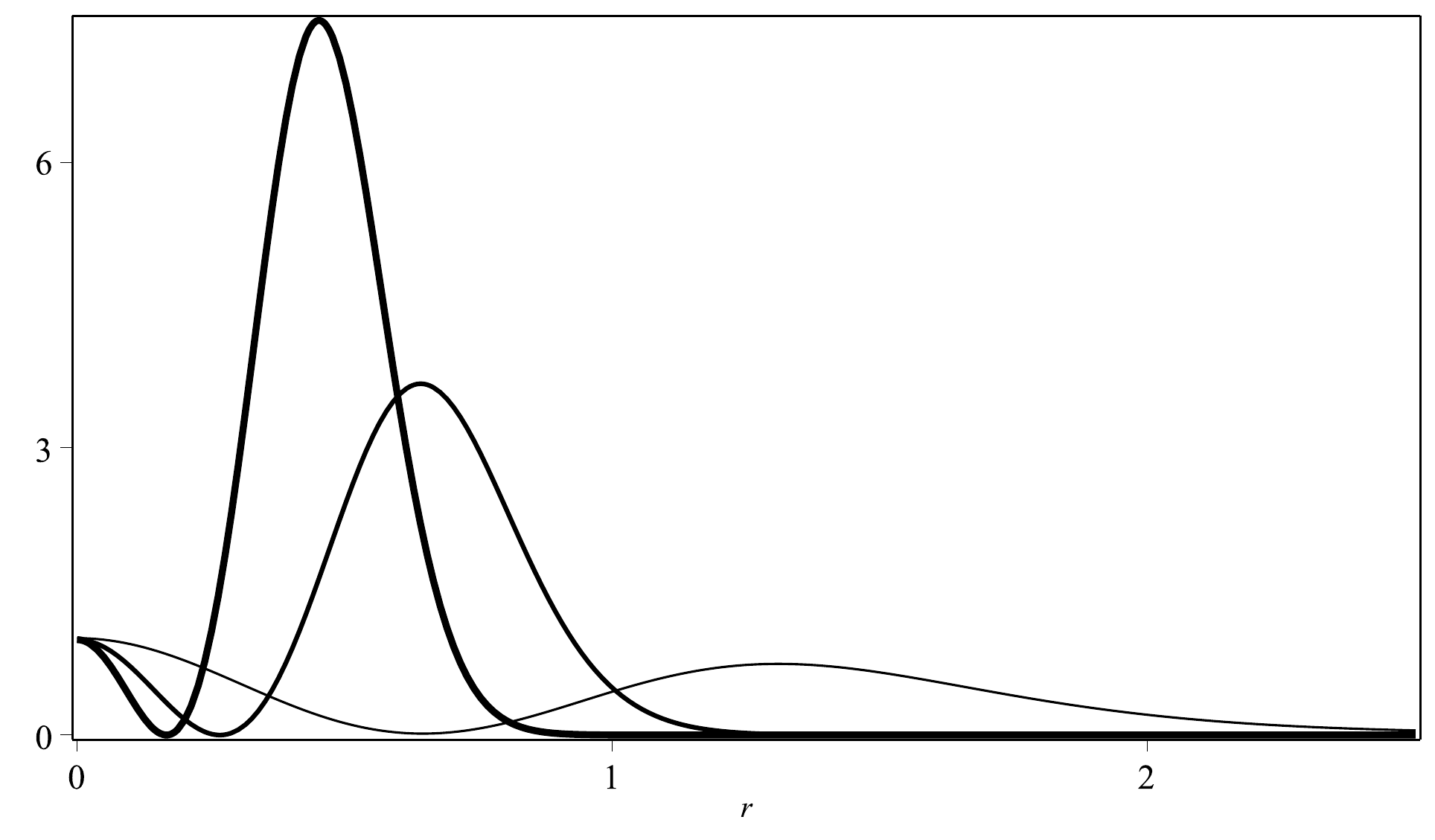}
\includegraphics[width=5.2cm]{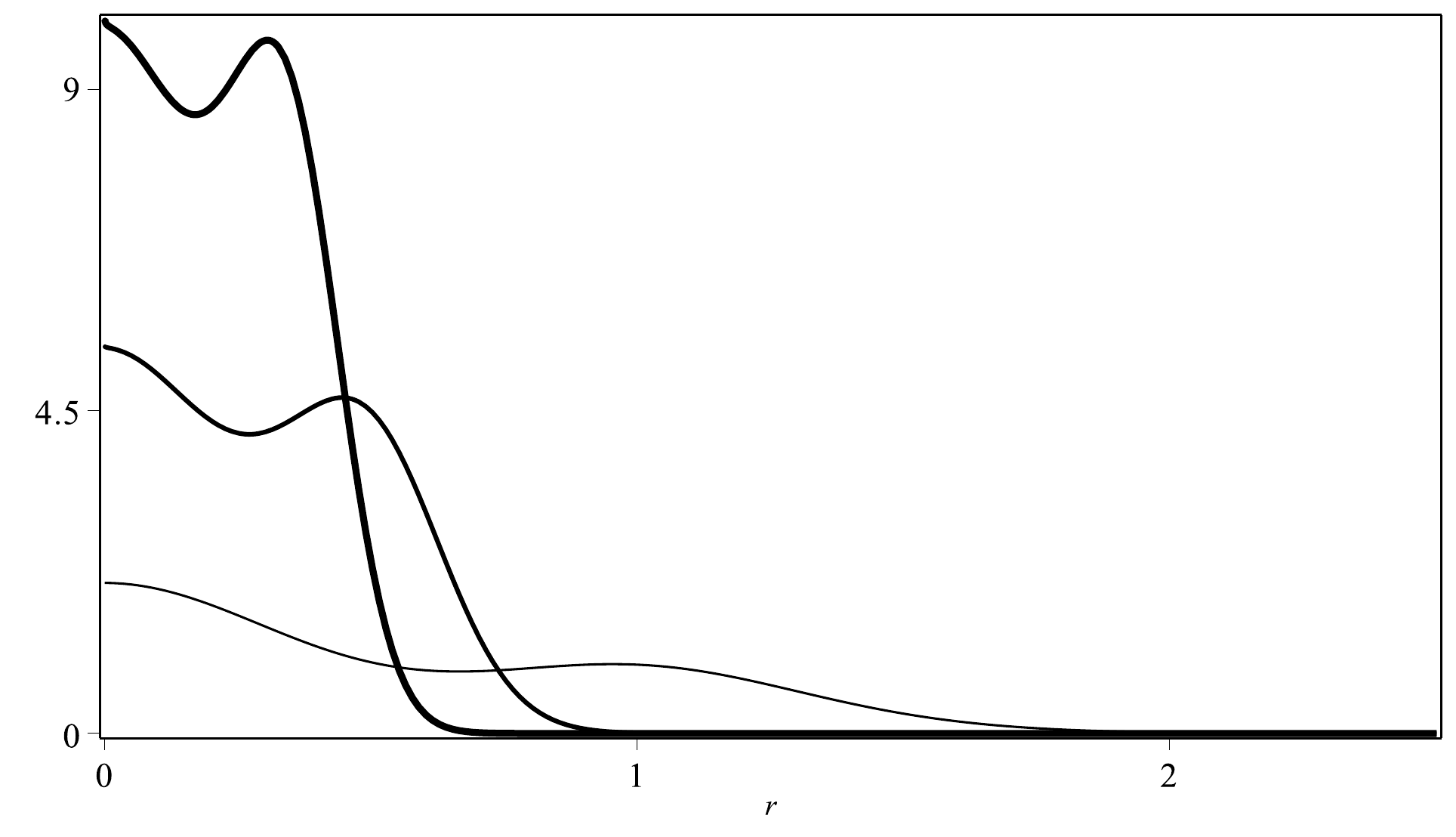}
\caption{The solutions $a(r)$ (descending lines) and $g(r)$ (ascending lines) of Eqs.~\eqref{fohiddenno} (top), the magnetic field (middle) and the energy density (bottom) of the visible sector for $q,w=1$ and $\beta=4, 20$ and $50$. The thickness of the lines increase with $\beta$.}
\label{fig1}
\end{figure}

In Fig.~\ref{fig1}, we plot the solutions of the above equations, the visible magnetic field \eqref{B} and the energy density \eqref{rhoans} for some values of $\beta$ and $q,w=1$. As $\beta$ increases, the solutions, the magnetic field and the energy density shrink. The solution $a(r)$, in particular, presents a region in which it is approximately uniform, as expected. This region is very wide for $\beta\approx1$ and tends to shrink and to become closer to the origin as we increase $\beta$. Due to this behavior in the solution, the magnetic field present a splitting. This also happens, in a more subtle manner, with the energy density. Therefore, the presence of the hidden sector generates an internal structure to the vortex engendered by the visible sector of the model. It is worth commenting that these features do not affect the magnetic flux and the energy of the vortex, which remain unchanged, given by Eqs.~\eqref{mflux} and \eqref{eb}.

\begin{figure}[t!]
\centering
\includegraphics[width=4.2cm]{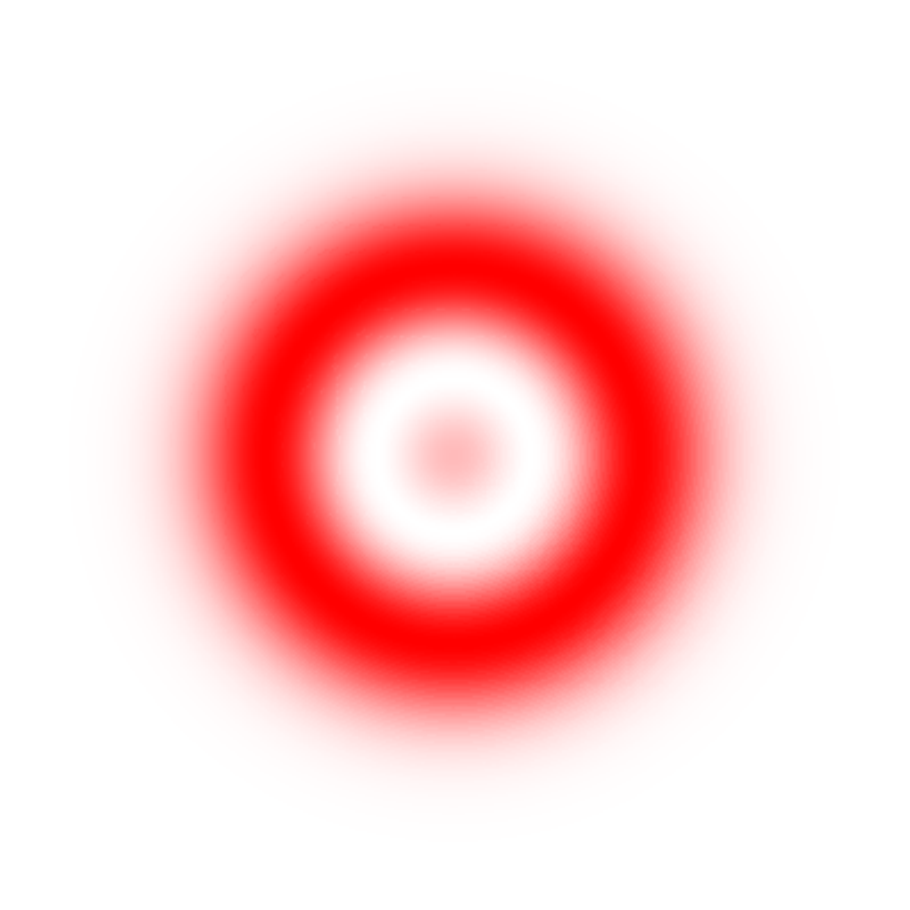}
\includegraphics[width=4.2cm]{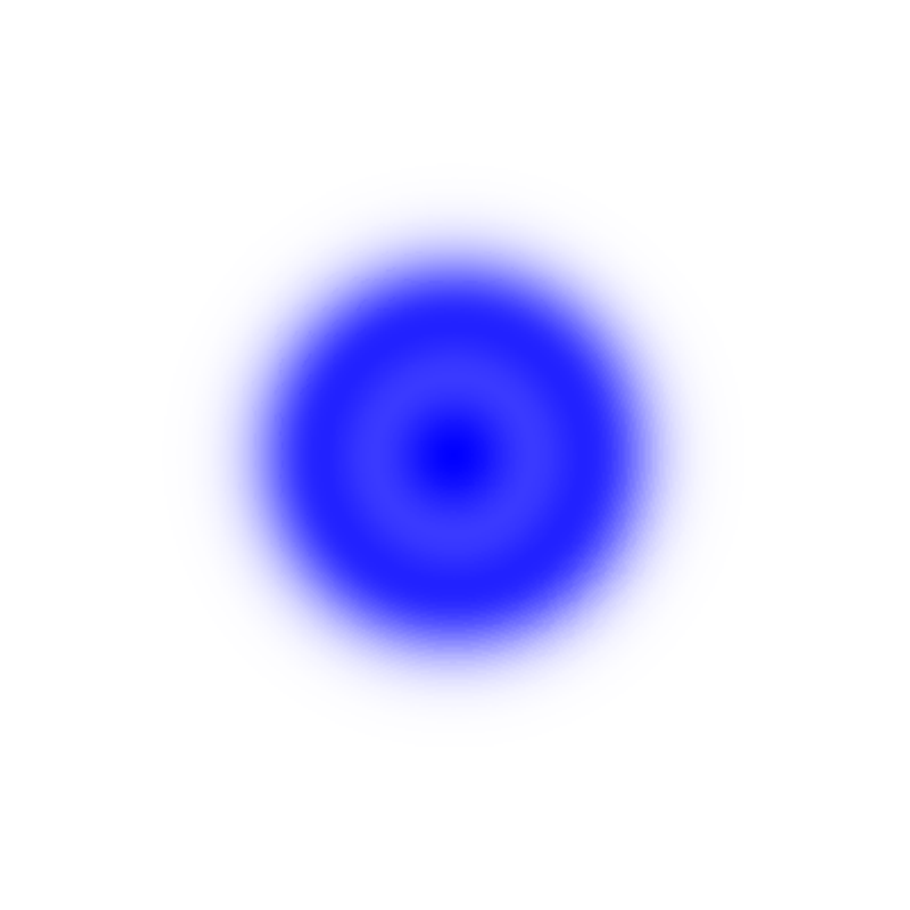}
\caption{The magnetic field (left) and the energy density (right) of the solutions of Eqs.~\eqref{fohiddenno} for the visible sector, with $q,w=1$ and $\beta=20$. The darkness of the colors is directly related to the intensity of the quantities.}
\label{fig2}
\end{figure}

The presence of a internal structure in the quantities related to the visible sector motivated us to plot the magnetic field and the energy density in the plane. It can be seen in Fig.~\ref{fig2}.

\subsection{Second example}
We now make a modification in the magnetic permeability of the hidden sector and suggest that it is driven by the function
\be
Q(|\chi|) =  \frac{w^2}{2\left(w^2-{|\chi|^2}\right)}.
\ee
According to the boundary conditions in Eq.~\eqref{bc}, $|\chi|\in[0,w]$; thus, $Q(|\chi|)$ is non negative where the solution exists. From Eq.~\eqref{fohidden}, we get the following first order equations for the hidden sector
\bes\label{fo2}
\bal
h^\prime &= \frac{ch}{r}, \\
-\frac{c^\prime}{qr} &= \frac{2q}{w^2} \left(w^2-{h^2}\right)^2.
\eal
\ees
In this case, we have been able to find their analytical solutions. They are given by
\be\label{solhidden2}
c(r) = \frac{1}{1+ (qwr)^2} \quad\text{and}\quad h(r) = \frac{qw^2r}{\sqrt{1+ (qwr)^2}}.
\ee
The above expression for $c(r)$ allows us to calculate the magnetic field associated to the hidden gauge field from Eq.~\eqref{B}, which is
\be\label{bhidden2}
\mathcal{B} = \frac{2qw^2}{\left(1+(qwr)^2\right)^2}.
\ee
The energy density of the hidden sector in Eq.~\eqref{rhohidden} may be also found explicitly
\be\label{rhohidden2}
\rho_{hidden} = \frac{2q^2w^4\left(3+(qwr)^2\right)}{\left(1+(qwr)^2\right)^4}.
\ee
Since the profiles of the functions in Eqs.~\eqref{solhidden2}-\eqref{rhohidden2} are very similar to the ones that appear in the well known Nielsen-Olesen case, they are not depicted here, although they are found analytically and easy to display.

\begin{figure}[t!]
\centering
\includegraphics[width=5.2cm]{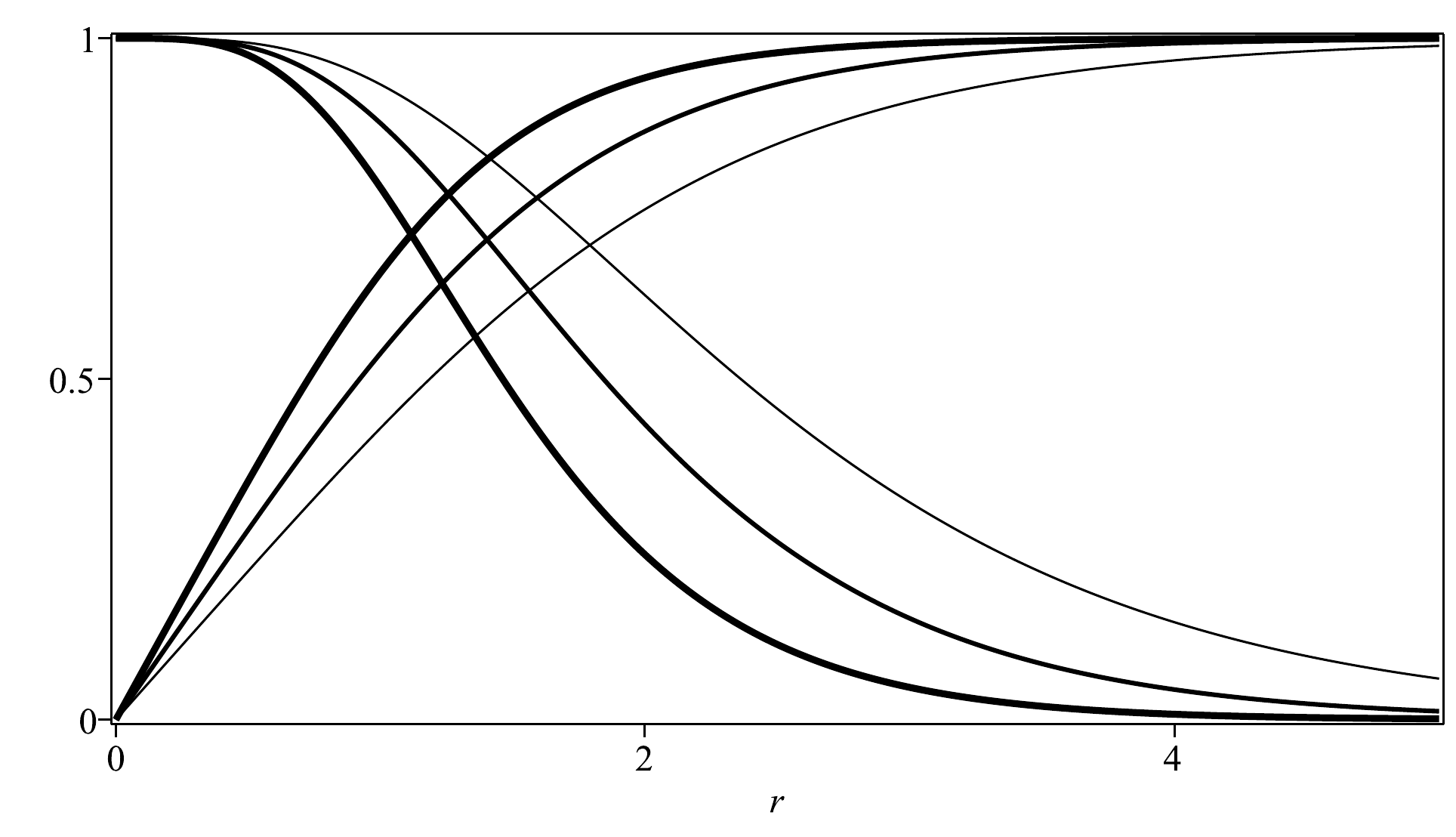}
\includegraphics[width=5.2cm]{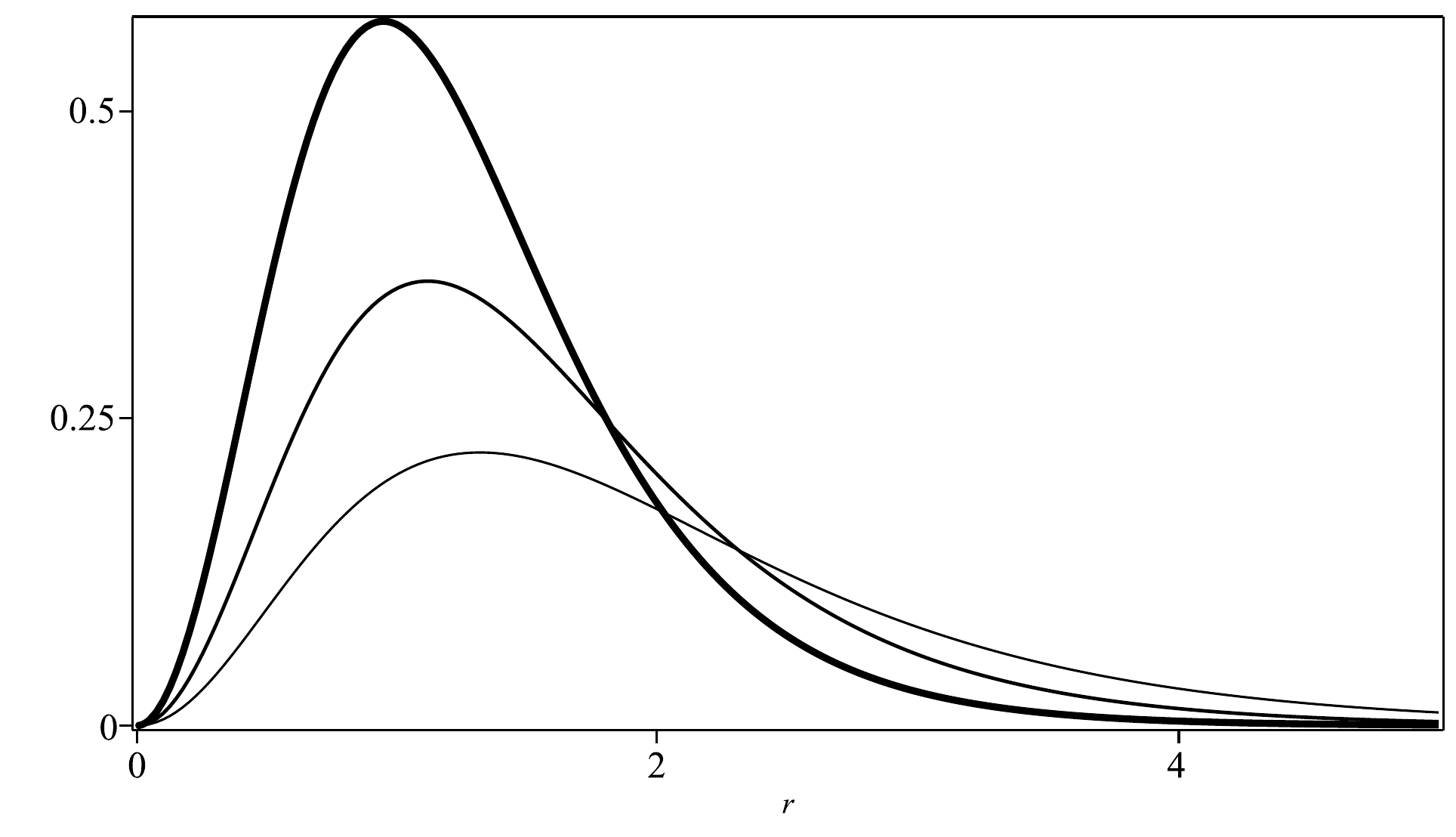}
\includegraphics[width=5.2cm]{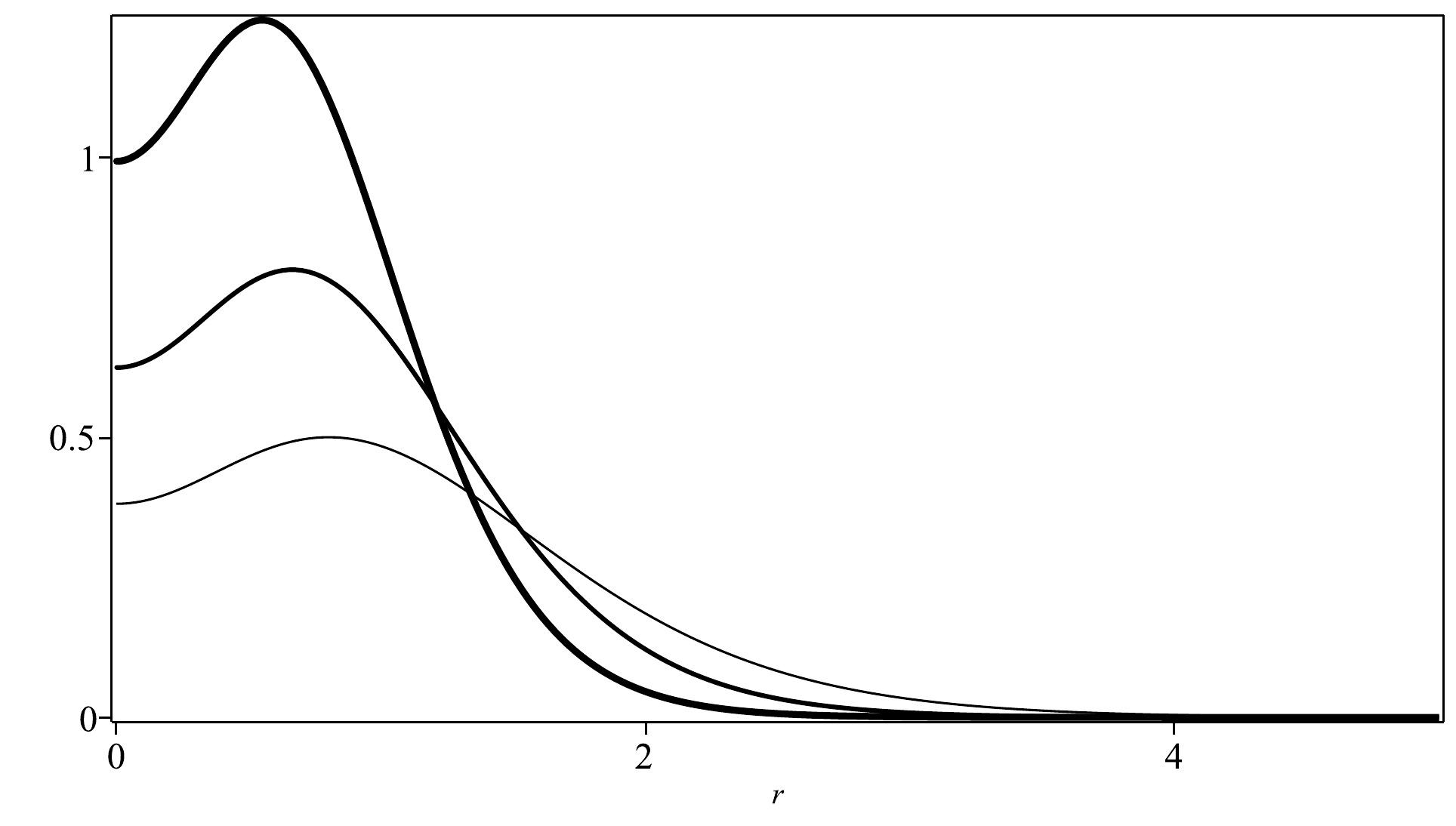}
\caption{The solutions $a(r)$ (descending lines) and $g(r)$ (ascending lines) of Eqs.~\eqref{fo2} (top), the magnetic field (middle) and the energy density (bottom) of the visible sector for $q,w=1$ and $\beta=0.5, 1$ and $2$. The thickness of the lines increases with $\beta$.}
\label{fig3}
\end{figure}

We then investigate how the solutions \eqref{solhidden2} modify the visible sector by considering its magnetic permeability to be governed by the function
\be
P(|\chi|) = \frac{w^2}{\beta|\chi|^2},
\ee
with $\beta$ real and positive. By doing so, the potential in Eq.~\eqref{potbogo} can be written as
\be
V(|\vphi|,|\chi|) =  \frac{\beta}{2}\! \left(1-|\vphi|^2\right)^2\!|\chi|^2 \!+\! \frac{q^2}{w^2}\!\left(w^2-|\chi|^2\right)^3.
\ee
In this case, the first order equations for the visible sector are
\bes\label{fo2}
\bal
g^\prime &= \frac{ag}{r}, \\
-\frac{a^\prime}{r} &= \beta\left(1-g^2\right)\frac{q^2w^2r^2}{1+q^2w^2r^2}.
\eal
\ees
The solutions of the above equations, the visible magnetic field \eqref{B} and energy density \eqref{rhoans} are depicted in Fig.~\ref{fig3}. Here, we highlight that the presence of the analytical solutions in Eq.~\eqref{solhidden2} allows us to show that the derivative of the solution $a(r)$ vanish at the origin. This makes the solution $a(r)$ approximately uniform in the neighborhood of this point. So, the solutions $a(r)$ present a plateau around the origin that leads to a hole in the magnetic field around the center of the vortex. This behavior also appear in the energy density, although in a subtle manner. As one can see here, the parameter $\beta$ is associated to the width of the solution and to the deepness of the valley around the origin in the magnetic field.

\begin{figure}[h!]
\centering
\includegraphics[width=4.2cm]{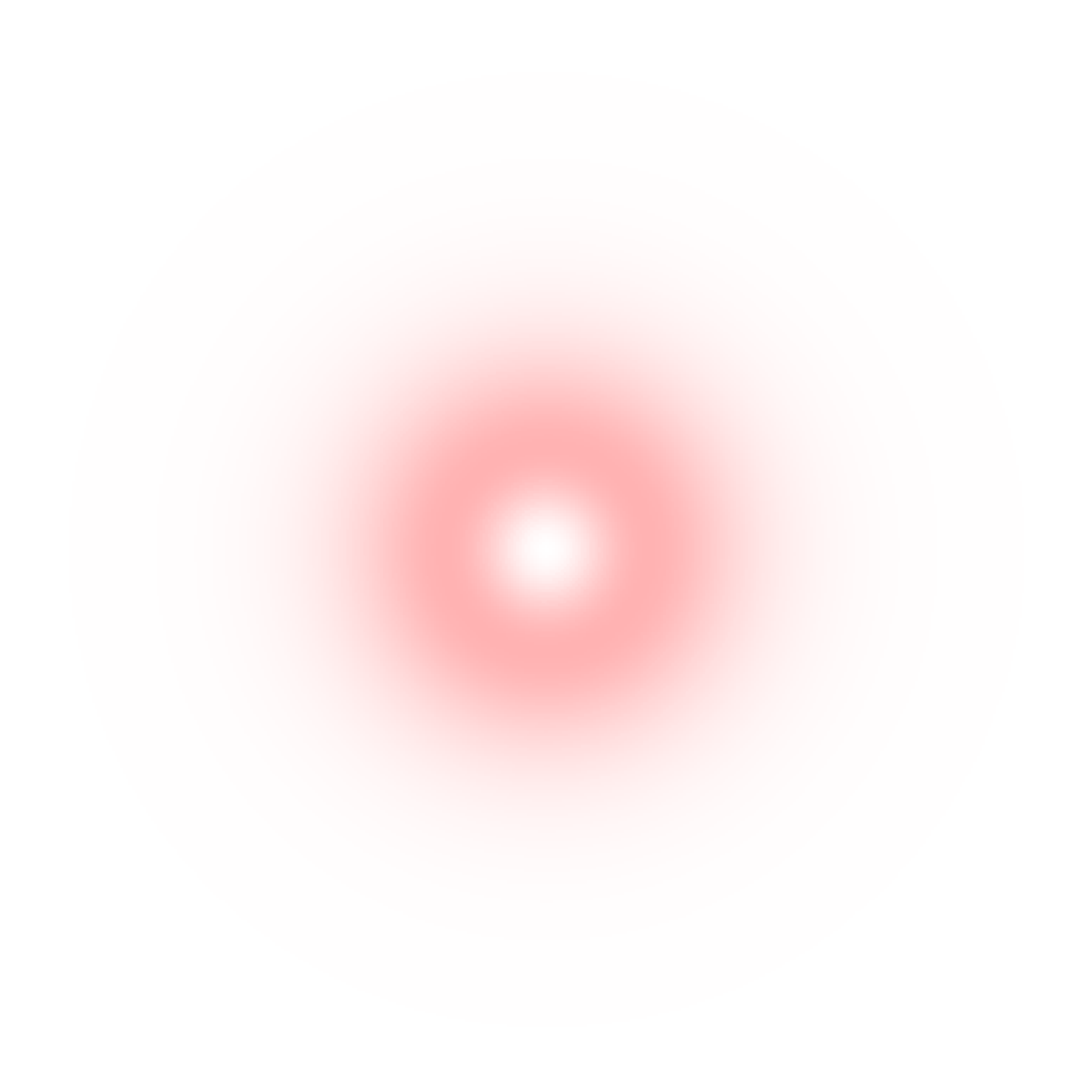}
\includegraphics[width=4.2cm]{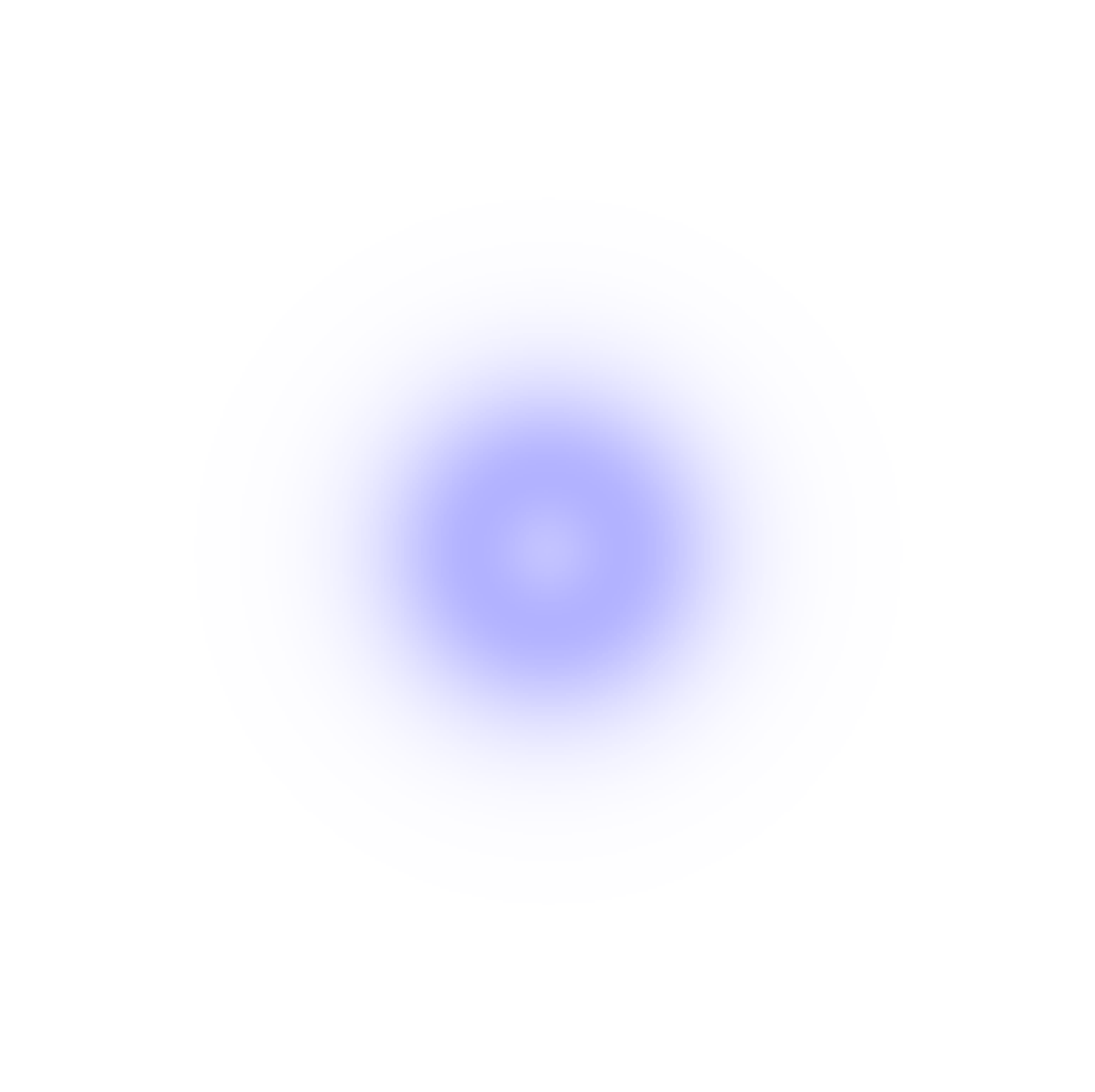}
\caption{The magnetic field (left) and the energy density (right) of the solutions of Eqs.~\eqref{fo2} for the visible sector, with $q,w=1$ and $\beta=2$. The darkness of the colors is as in Fig.~\ref{fig2}.}
\label{fig4}
\end{figure}

This model engenders a distinct behavior from the previous one in a qualitative manner, presenting a hole in magnetic field near the origin. The new features supported by our models motivated us to depict the magnetic field and the energy density for the solutions of Eqs.~\eqref{fo2} of the visible sector in the plane. It can be seen in Fig.~\ref{fig4}. As it was mentioned before, the presence of a hole is stronger in the magnetic field than in the energy density.

\section{Conclusions}\label{secconc}
In this work, we studied vortex structures in generalized Maxwell-Higgs models with visible and hidden sectors. The models considered in this work  contain two additional functions, $P(|\chi|)$ and $Q(|\chi|)$, that control the magnetic permeability of the visible and hidden sectors, respectively. The interaction between the two sectors are controlled by the hidden scalar field which appears in the function $P(|\chi|)$.

We have chosen an specific form of the potential, which allowed for the Bogomol'nyi procedure to work out, and for the presence of first order differential equations that solve the equations of motion. The procedure has shown that the first order equations of the hidden sector decouple from the visible one and could be solved independently. In this sense, the hidden sector acts as a source for the solutions in the visible sector. By taking specific forms for the magnetic permeabilities, we have found that the hidden charged scalar field affects the visible vortex configuration, generating an internal structure to it. In the first example, the magnetic field present an apparent valley outside the origin that seems to connect two separated structures. The effect is less evident in the energy density, although it is also there. In the second example, the vortex presents a magnetic field with a hole around the origin. Surprisingly, the energies and fluxes of the hidden and visible vortices are fixed by the boundary conditions, and they are independent from each other. Moreover, they do not depend on the specific form we choose for the magnetic permeabilities to construct the system. As a particularly interesting result, we could find a specific function $Q(|\chi|)$ which allowed us to describe the vortex solution of the hidden sector, and the corresponding magnetic field and energy density, analytically. 

Here we have discarded the coupling between the two electromagnetic strength tensors, so an issue to be further examined would be to add this kind of coupling. Another possibility is to generalize the covariant derivative terms, as suggested before in \cite{anavortex}. Since the model examined above attains the Bogomol'nyi bound, it appears to be the bosonic portion of a lager, supersymmetric theory, and this is presently under consideration, to see how supersymmetry is working to lead us to the first order equations.

We are also thinking of enlarging the model to accommodate other symmetries, such as the $SU(2)\times U(1)$ symmetry, to explore the presence of solutions within the non-Abelian context examined before in \cite{sup2}, and also the $SU(3)\times U(1)$ symmetry, to investigate color-magnetic structures in the dense quark matter scenario explored recently in \cite{cm}. The presence of non-Abelian symmetries makes the problem much harder, so the search for first order differential equations that solve the equations of motion is of current interest and would be welcome. 

Another possibility is to use the same $U(1)\times U(1)$ symmetry, but now changing the Maxwell dynamics of the gauge field in the hidden sector by the Chern-Simons one, to see how the Chern-Simons vortex may modify the Maxwell-Higgs vortex in the visible sector. We hope to report on some of these issues in the near future.

\acknowledgements{We would like to acknowledge the Brazilian agency CNPq for partial financial support. DB thanks support from grant 306614/2014-6, LL thanks support from grant 303824/2017-4, MAM thanks support from grant 155551/2018-3 and RM thanks support from grant 306504/2018-9.}


\end{document}